\documentclass[runningheads]{svmult}

\usepackage{makeidx}   
\usepackage{graphicx}  
\usepackage{subeqnar}  
\usepackage{physprbb}  
\makeindex             

\def\be{\begin{equation}}
\def\ee{\end{equation}}
\def\simless{\mathbin{\lower 3pt\hbox
   {$\rlap{\raise 5pt\hbox{$\char'074$}}\mathchar"7218$}}} 
\def\simgreat{\mathbin{\lower 3pt\hbox
   {$\rlap{\raise 5pt\hbox{$\char'076$}}\mathchar"7218$}}}

\begin{document}

\title*{Phase Diagram for Spinning and Accreting Neutron Stars}

\toctitle{Phase Diagram for Spinning and Accreting Neutron Stars}

\titlerunning{Phase Diagram for Accreting Neutron Stars}

\author{David Blaschke \inst{1}
\and Hovik Grigorian \inst{2} 
\and Gevorg Poghosyan \inst{2} }

\authorrunning{D. Blaschke et al.}

\institute{Department of Physics,
University of Rostock, D-18051 Rostock, Germany
\and
Department of Physics, 
Yerevan State University, Alex Manoogian 1, 375025 Yerevan, Armenia
}

\maketitle              

\begin{abstract}
Neutron star configurations are considered as thermodynamical systems for which
a phase diagram in the angular velocity ($\Omega$) - baryon 
number ($N$) plane is obtained with a dividing line $N_{\rm crit}(\Omega)$ 
for quark core configurations.
Trajectories of neutron star evolution in this diagram are studied for 
different scenarios defined by the external torque acting on 
the star due to radiation and/or mass accretion.
They show a characteristic change in the rotational kinematics when the star
enters the quark core regime. 
For isolated pulsars the braking index signal for deconfinement has been 
studied in its dependence on the mass of the star.
Model calculations of the spin evolution of accreting low-mass X-ray binaries 
in the phase diagram have been performed for different values of the initial 
magnetic field, its decay time as well as initial mass and mass accretion rate.
Population clustering of these objects at the line $N_{\rm crit}(\Omega)$ in
the phase diagram is suggested as an observable signal 
for the deconfinement phase transition if it exists for spinnning and 
accreting neutron stars.
\end{abstract}

\section{Neutron stars as thermodynamical systems}

Quantum Chromodynamics (QCD) as the fundamental theory for strongly interacting
matter predicts a deconfined state of quarks and gluons under 
conditions of sufficiently high temperatures and/or densities which occur, 
e.g., in heavy-ion collisions, \index{Heavy-ion collisions}
a few microseconds after the Big Bang or in the cores of pulsars. 
The unambiguous detection of the phase transition from hadronic to quark matter
(or vice-versa) has been a challenge to particle and astrophysics 
over the past two decades \cite{qm99,bkr}. 
While the diagnostics of a phase transition in experiments with heavy-ion beams
faces the problems of strong nonequilibrium and finite size, the dense
matter in a compact star forms a macroscopic system in thermal and chemical 
equilibrium for which signals of a phase transition shall be more pronounced.

Such signals have been suggested in the form of characteristic changes of 
observables such as the surface temperature \cite{cool}, 
brightness \cite{magnetar}, pulse timing \cite{frido} and rotational 
mode instabilities \cite{madsen} during the evolution of the compact object.
In particular the pulse timing signal has attracted much interest since it is 
due to changes in the kinematics of rotation. Thus it could be used 
not only to detect the occurrence but also to determine the size of the quark 
core from the magnitude of the braking index deviation from the magnetic 
dipole value \cite{cgpb}. 
Besides of the isolated pulsars, one can consider also the accreting 
compact stars in low-mass X-ray binaries (LMXBs) as objects from which we can 
expect signals of a deconfinement transition in their interior 
\cite{cgpb,fridonew,accmag}. 
The observation of quasiperiodic brightness oscillations (QPOs) \cite{klis} 
for some LMXBs has lead to very stringent constraints for masses and radii 
\cite{MLP} which according to \cite{strange,bombachi} could even favour strange quark 
matter interiors over hadronic ones for these objects.
Due to the mass accretion flow these systems are candidates for the 
formation of the most massive compact stars from which we expect to observe 
signals of the transition to either quark core stars, to a third family of 
stars \cite{twins} or to black holes.  
Each compact star configuratrion can be identified with a thermodynamical 
system characterized by the total baryon number (or mass), temperature, 
spin frequency and magnetic field as thermodynamical variables. 

Since the evolutionary processes for the compact objects accompanying the 
structural changes are slow enough we will consider here the case of rigid 
rotation only and restrict ourselves to the degenerate systems at $T=0$.
The magnetic and thermal evolution of the neutron stars we will consider as 
decoupled from the mechanical evolution. 

\begin{figure}
\includegraphics[width=1.25\textwidth]{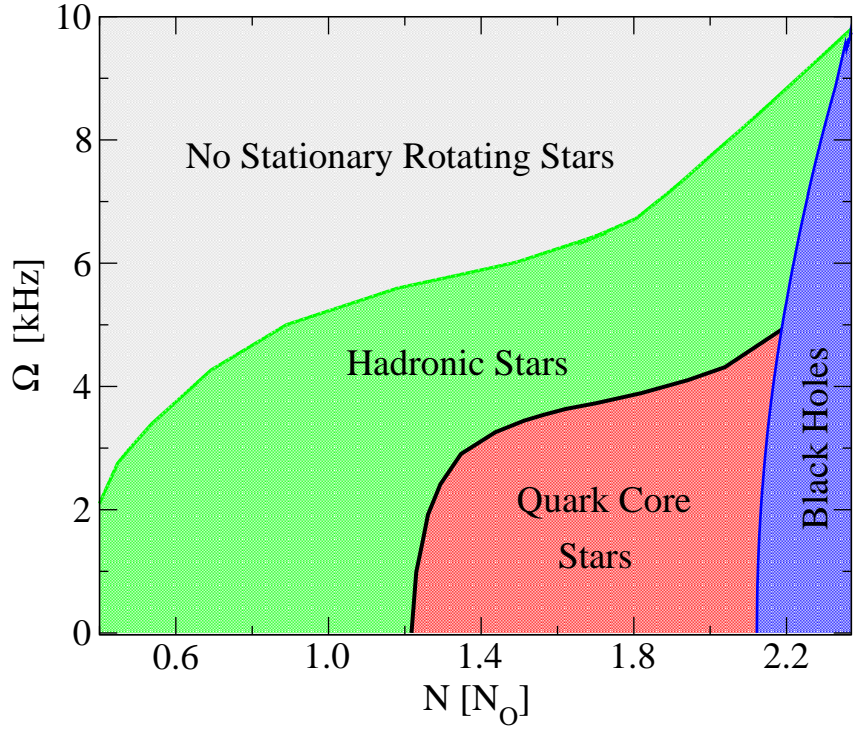}
\caption[]{Sketch of a phase diagram for spinning neutron stars with 
deconfinement phase transition. }
\label{phases}
\end{figure}

In this approximation we can introduce a classification of isolated and 
accreting compact stars in the plane of their angular frequency $\Omega$ and 
mass (baryon number $N$) which we will call {\it phase diagram}, 
\index{Phase diagram!for rotating stars}
see Fig. \ref{phases}. 

Each point in this phase diagram corresponds to a mechanical state of a 
neutron star.
Mechanical equilibrium of thermal pressure with gravitational and centrifugal 
forces leads to a stationary distribution of matter inside the configuration.
This distribution is determined by the central density and rotational 
frequency.
Thermal and chemical equilibrium are described by an equation of state which
determines the structure and composition of the compact star configuration.  

The requirements of stability restrict the region of stability of the 
quasistationary rotating star configurations. From the right hand side of the 
phase diagram there is a line which separates a region of stars which are 
collapsed to black holes (BH) from the stable ones. At high baryon numbers 
beyond this line gravitational forces dominate over pressure and centrifugal 
forces of matter.  

From the top of the diagram the region of stable star configurations is 
separated by the Keplerian frequencies $\Omega_K(N)$ from that where  mass 
shedding under the centrifugal forces does not allow stationary rotating 
objects.

Inside the region of stationary rotators one can distinguish two types of 
stationary rotating stars: those with quark matter cores (QCSs) and hadronic 
stars (HS) \cite{phdiag}.
The region of the QCSs is expected to be located at the bottom right of the 
phase diagram in the region of the most massive and slowly rotating compact 
stars.
The critical line $N_{\rm crit}(\Omega)$ which separates the QCS region from 
that of HSs is correlated with the local maxima of the moment of inertia with 
respect to changes of the baryon number at given angular velocity $\Omega$ due 
to the change of the internal structure of the compact object at the 
deconfinement transition \index{Phase diagram!critical line}.

Using the analogy with the phase diagram of a conventional thermodynamical 
system we can consider trajectories in this diagram as quasistationary 
evolutionary processes. 
  
Since these processes for neutron stars have characteristic time scales 
($\approx 10^6 \div 10^8$ yr) 
much longer than those of typical observers it is almost impossible to trace 
these trajectories directly. A possible strategy for measuring evolutionary 
tracks will be to perform a statistics  of the population of the different 
regions in the phase diagram. \index{Neutron star!spin!evolution}

Our aim is to investigate the conditions under which the passage of the phase 
border leads to observable significant clustering in the populations. 

We will provide criteria under which a particular astrophysical 
scenario with spin evolution could be qualified to signal a deconfinement 
transition. \index{Phase transition!deconfinement!signals}

\section{Phase diagram for a stationary rotating star model}

\subsection{Equation of state with deconfinement transition}
\index{Quark!deconfinement}
Since our focus is on the elucidation of qualitative features of signals 
from the deconfinement transition in the pulsar timing we will use a generic 
form of an equation of state (EoS) with such a transition \cite{cgpb} which 
is not excluded by the mass and radius constraints derived from QPOs.
In our case as well as in most of the approaches to quark deconfinement in 
neutron star matter a standard two-phase description of
the equation of state is applied where the hadronic phase and the quark
matter phase are modelled separately. 
The ambiguity in the choice of the bag constant for the description of 
the quark matter phase can be removed by a derivation of this quantity 
\cite{b+99,bt00} from a dynamical confining approach \cite{bbkr}.
The resulting EoS is obtained by imposing Gibbs' conditions for phase
equilibrium with the constraints of globally conserved baryon number and 
electric charge \cite{glenv,gbook,fwbook}. 

Since nuclear forces are not fundamental, our knowledge about the equation of 
state for nuclear and neutron star matter at high densities is not very 
precise. There is not yet a description of the equation of state for strongly 
interacting matter on the fundamental level in terms of quark and gluon 
degrees of freedom where nucleons and mesons appear as composite structures.
It is one of the goals of nuclear astrophysics and of neutron star 
physics to use pulsars and other compact objects as laboratories for studies 
of the nuclear forces and the phases of nuclear matter \cite{gbook,fwbook}.
The existence of several models for the equation of state for dense stellar  
matter allows a variability in the phase structure of neutron stars 
\cite{baldo,prakashv}, therefore the phase diagram (Fig. \ref{phases}) remains
robust only qualitativly.

For the detailed introduction of the phase diagram and a quantitative
analysis using thermodynamical
methods, we will employ a particular EoS model \cite{cgpb}
which is characterized by a relatively hard hadronic matter part.

\subsection{Configurations of rotating stars}

In our model calculations we assume quasistationary evolution
with negligible convection and without differential rotation
which is justified when both the  \index{Neutron star!rotating}
mass load onto the star and the transfer of the angular momentum are 
sufficiently slow processes.

For our treatment of rotation within general relativity we employ a 
perturbation expansion with respect to the ratio of the rotational and 
gravitational energies for the homogeneous Newtonian spherical 
rotator with the mass density $\rho(0)$ equal to the central density,
$E_{\rm rot}/E_{\rm grav}=(\Omega/\bar{\Omega})^2 $, 
where  $\bar{\Omega}^2 = 4\pi G \rho(0)$. 
This ratio is a small parameter, less than one up to the mass shedding 
limit \cite{cgpb}. \index{Neutron star!rotating!perturbation expansion}
\index{Neutron star!rotating!slowly}

The general form of the expansion  allows us to describe
the metric coefficients and the distributions of pressure, energy density
and hydrodynamical enthalpy in the following form
\begin{eqnarray}  \label{euler1}
X(r,\theta;\Omega) &=&X^{(0)}(r)+(\Omega/\bar{\Omega})^2 X^{(2)}(r,\theta
)+O(\Omega^4)~, 
\end{eqnarray}
where $X$ stands for one of the above mentioned quantities \cite{cgpb}.
The series expansion allows one to transform the Einstein equations into a
coupled set of differential equations for the coefficient functions defined in 
(\ref{euler1}), which can be solved by recursion.
The static solutions, obeying the Oppenheimer-Volkoff equations, are contained 
in this expansion for the case $\Omega=0$ when only the functions with 
superscript ($0$) remain. 
The other terms are corrections due to the rotation.
We truncate higher order terms $\sim O(\Omega^4)$ in this expansion and neglect
the change of the frame dragging frequency, which appears at $O(\Omega^3)$.
For a more detailed description of the method and analytic results in the
integral representation of the moment of inertia  we refer to 
\cite{cgpb} and to works of Hartle and Thorne\cite{hartle,thorne}, 
Sedrakian and Chubarian \cite{chubarian,sedrakian}. 

\begin{figure}[bht]
\centerline{\includegraphics[height=.98\textwidth,angle=-90]{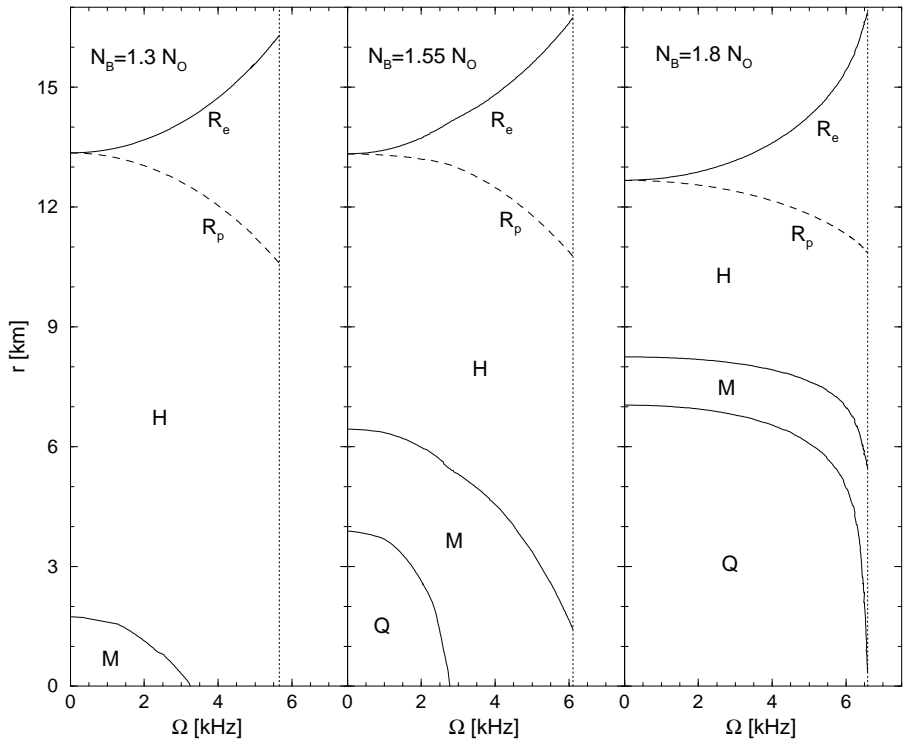}}
\caption
{Phase structure of rotating hybrid stars in equatorial 
direction in dependence of the angular velocity $\Omega$ for stars 
with different total baryon number: $N_B/N_{\odot}=1.3, 1.55, 1.8$.
\label{fig:struct}}
\end{figure}

In Fig. \ref{fig:struct} we show the critical regions of the phase transition 
in the inner structure \index{Neutron star!structure} of the star configuration as well as the
equatorial and polar radii in the plane of angular velocity $\Omega$
versus distance from the center of the star \index{Neutron star!hybrid}. 
It is obvious that with the increase of the
angular velocity the star is deformating its shape. 
The maximal excentricities of the configurations with $N_B=1.3~N_\odot$,  
$N_B=1.55~N_\odot$ and $N_B=1.8~N_\odot$ are 
$\epsilon(\Omega_{\rm max})=0.7603$, 
$\epsilon(\Omega_{\rm max})=0.7655$ and 
$\epsilon(\Omega_{\rm max})=0.7659$, respectively. 
Due to the changes of the central density the quark core could disappear above
a critical angular velocity. 
 
\begin{figure}[bht]
\begin{center}
\includegraphics[width=1.0\textwidth]{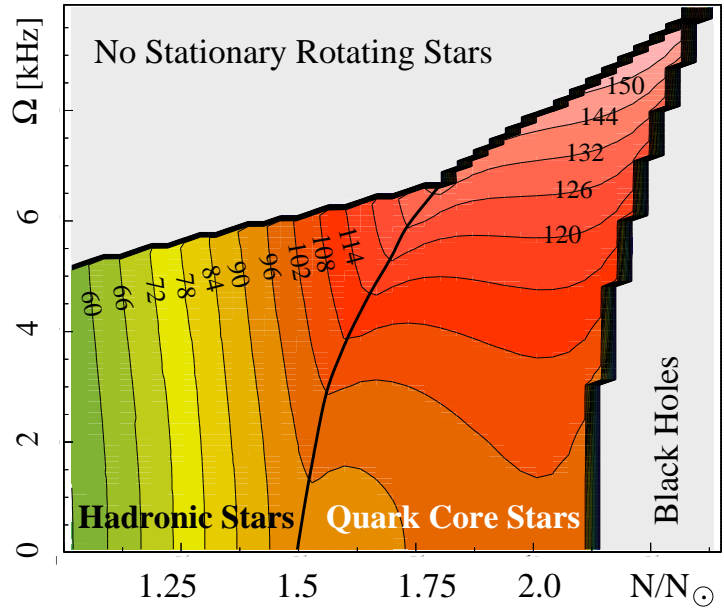}
\end{center}
\caption[]{Phase diagram for configurations of rotating compact objects in 
the plane of angular velocity $\Omega$ and mass (baryon number $N$).
Contour lines show the values of the moment of inertia in M$_\odot$km$^2$.
The line $N_{\rm crit}(\Omega)$ which separates hadronic from quark core stars
corresponds the set of configurations with a central density equal to the 
critical density for the occurence of a pure quark matter phase.
\index{Phase transition!and moment of inertia}}
\label{fig:phdiag}
\end{figure}

It is the aim of the present paper to investigate the conditions for a 
verification of the existence of the critical line $N_{\rm crit}(\Omega)$
by observation.  
We will show evidence that in principle such a measurement is possible since
this deconfinement transition line corresponds to a maximum of the moment of 
inertia, which is the key quantity for the rotational behavior of compact 
objects, see Fig. \ref{fig:phdiag}.

In the case of rigid rotation the moment of inertia is defined by
\index{Neutron star!moment of inertia}
\begin{equation}
\label{momi}
I(\Omega, N) = J(\Omega, N)/\Omega~,
\end{equation}
where the angular momentum $J(\Omega, N)$ of the star can be expressed in 
invariant form as
\begin{equation}
J(\Omega, N)=\int T_\phi ^t\sqrt{-g}dV~,  \label{moment}
\end{equation}
with $T_\phi ^t$ being the nondiagonal element of the energy momentum
tensor,  $\sqrt{-g}dV$  the invariant volume and
$g=\det||g_{\mu\nu}||$ the determinant of the metric tensor. 
We assume that the superdense compact object rotates stationary 
as a rigid body, so that for a given time-interval both the angular velocity 
as well as the baryon number can be
considered as global parameters of the theory.
The result of our calculations for the moment of inertia (\ref{momi}) can be 
cast into the form
\begin{equation}
I= I^{(0)}+\sum_{\alpha}\Delta I_{\alpha},
\end{equation}
where  $I^{(0)}$ is the moment of inertia of the static configuration
with the same central density and $\Delta I_{\alpha}$ 
stands for contributions to the  moment of inertia from different rotational
effects which are labeled by $\alpha$: matter redistribution, 
shape deformation, and changes in the centrifugal forces and the gravitational 
field \cite{cgpb}. 

In Fig.  \ref{fig:phdiag} we show the resulting phase diagram for
compact star configurations which exhibits four regions:  (i) the
region above the maximum frequency $\Omega_{\rm K}(N)$ where no
stationary rotating configurations are found, (ii) the region of black
holes for baryon numbers exceeding the maximum value $N_{\rm
max}(\Omega)$, and the region of stable compact stars which is
subdivided by the critical line $N_{\rm crit}(\Omega)$ into a region
of (iii) quark core stars and another one of (iv) hadronic stars,
respectively.  The numerical values for the critical lines are model
dependent.  For this particular model EoS due to the hardness of the
hadronic branch (linear Walecka model \cite{gbook}) there is a maximum
value of the baryonic mass on the critical line $N_{\rm
crit}(\Omega_k)=1.8 N_{\odot}$, such that for stars more massive
than that one all stable rotating configurations have to have a quark core.
This property can be seen from the dependence of the phase structure
of the star on angular velocity in Fig. \ref{fig:struct}.  For the
whole interval of possible frequencies in the case of $N=1.8
N_{\odot}$ the quark core radius remains approximately unchanged: 
$R_{core} \sim 7$ km.

\section{Evolution scenarios with phase transitions}
\index{Phase transtition!in neutron star}
We want to explain why the occurence of a sharply peaked maximum for the  
moment of inertia in the 
$\Omega - N$ plane entails observational consequences for the angular velocity
evolution of rotating compact objects. 
The basic formula which governs the rotational dynamics is
\begin{equation}
\dot{\Omega}=
\frac{K(N,\Omega)}
{I(N,\Omega)+\Omega\left({\partial I(N,\Omega)}/{\partial \Omega}\right)_N}~,
\end{equation}
where $K=K_{\rm int}+K_{\rm ext}$ is the net torque acting on the star due to 
internal and external forces. 
The internal torque is given by
$K_{\rm int}(N,\Omega)=-\Omega\dot N 
\left({\partial I(N,\Omega)}/{\partial N}\right)_\Omega$~, the external one 
can be subdivided into an accretion 
and a radiation term $K_{\rm ext}=K_{\rm acc} + K_{\rm rad}$. 
The first one is due to all processes which change the baryon number, 
$K_{\rm acc}=\dot N~dJ/dN$ and the second one 
contains all processes which do not. 
For the example of magnetic dipole and/or gravitational wave radiation it can 
be described by a power law $K_{\rm rad}=\beta\Omega^{n}$, see 
\cite{ghoshlamb,shapirov}.\index{Neutron star!spin!evolution!dynamics}

\subsection{Spin-down scenario for isolated pulsars}

The simple case of the spindown evolution of  
isolated (non-accreting, $\dot N=0$) pulsars due to magnetic dipole radiation 
would be described by vertical lines in Fig. \ref{phases}.
These objects can undergo a 
deconfinement transition if the baryon number lies within the interval
$N_{\rm min}< N < N_{\rm max}$, 
where for our model EoS the endpoints of $N_{\rm crit}(\Omega)$ are
$N_{\rm min} = 1.49 ~N_\odot$ and $N_{\rm max} = 1.78 ~N_\odot$.
As it has been shown in \cite{cgpb}, the braking index 
$n(\Omega)$ changes its value from $n(\Omega)>3$ in the region (iii) to 
$n(\Omega)<3$ in (iv). This is the braking index signal for a 
deconfinement transition introduced by \cite{frido}. 

\begin{figure}[bht]
\includegraphics[width=1.0\textwidth,height=0.5\textheight,clip=0]{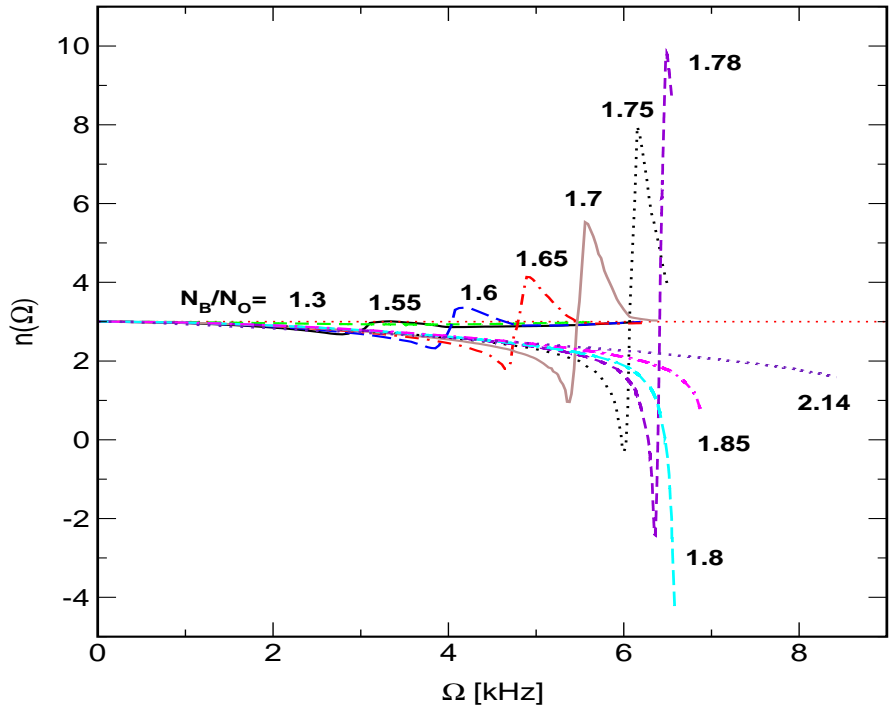}
\caption
{Braking index  due to dipole radiation from fastly rotating 
isolated pulsars as a function of the angular velocity.
The minima of $n(\Omega)$ indicate the appearance/ disappearance 
of quark matter cores.
\label{fig:brake}}
\end{figure}

In Fig. \ref{fig:brake} we display the result for the braking index $n(\Omega)$ 
\index{Braking index}
for a set of configurations with fixed total baryon numbers ranging from 
$N_B=1.55~N_\odot$ up to $N_B=1.9~N_\odot$, the region where during the 
spin-down evolution a quark matter core could occur for our model EOS.
We observe that only for configurations within the interval of total 
baryon numbers $1.4\le N_B/N_\odot\le 1.9$ a quark 
matter core occurs during the spin-down as a consequence of the increasing 
central density, see also Fig. \ref{fig:struct}, and the braking index shows 
variations. \index{Braking index!during phase transition}
The critical angular velocity $\Omega_{\rm crit}(N_B)$ for the appearance of 
a  quark matter core can be found from the minimum of the braking index $n(\Omega)$.
As can be seen from Fig. \ref{fig:brake}, all configurations with a quark 
matter core have braking indices $n(\Omega) < 3$ and braking indices 
significantly larger than 3 can be considered as precursors of the 
deconfinement transition. 
The magnitude of the jump in  $n(\Omega)$ during the transition to 
the quark core regime is a measure for the size of the quark core. 
It would be even sufficient to observe the maximum of the braking index 
$n_{\rm max}$ in order to infer not only the onset of deconfinement 
($\Omega_{\rm max}$) but also the size of the quark core to be 
developed during further spin-down from the maximum deviation 
$\delta n = n_{\rm max} - 3$ of the braking index.
For the model EOS we used a significant enhancement of the braking 
index does only occur for pulsars with periods $P<1.5$ ms (corresponding to 
$\Omega>4$ kHz) which have not yet been observed in nature.
Thus the signal seems to be a weak one for most of the possible candidate 
pulsars.
However, this statement is model dependent since, e.g., for the model EOS 
used by \cite{frido}, which includes the 
strangeness degree of freedom, a more dramatic signal at lower spin 
frequencies has been reported.
Therefore, a more complete investigation of the braking index for a set of 
realistic EOS should be performed.

\subsection{Scenarios with mass accretion}

All other possible trajectories
correspond to processes with variable baryon number (accretion).
In the phase of hadronic stars, $\dot \Omega$ first decreases as long as
the moment of inertia monotonously increases with $N$.  
When passing the critical line $N_{\rm crit}(\Omega)$ for the deconfinement 
transition, the moment of inertia starts decreasing and the internal torque 
term $K_{\rm int}$ changes sign. This leads to a narrow dip for 
$\dot \Omega (N)$ in the vicinity of this line. 
As a result, the phase diagram gets populated for 
$N \stackrel{<}{\sim} N_{\rm crit}(\Omega)$ and depopulated for 
$N \stackrel{>}{\sim} N_{\rm crit}(\Omega)$ up to the second maximum of 
$I(N, \Omega)$ close to the black-hole line $N_{\rm max}(\Omega)$. 
The resulting population clustering of compact stars
at the deconfinement transition line is suggested to 
emerge as a signal for the occurence of stars with quark matter cores.
In contrast to this scenario, in the case without a deconfinement transition, 
the moment of inertia could at best saturate before the transition to the 
black hole region and consequently $\dot \Omega$ would also saturate. 
This would entail a smooth population of the phase diagram without a 
pronounced structure.

The clearest scenario could be the evolution along lines of constant 
$\Omega$ in the phase diagram. 
These trajectories are associated with processes where the external and 
internal torques are balanced. A situation like this has been described, 
e.g. by \cite{bildsten} for accreting binaries emitting gravitational waves. 

\begin{figure}[htb]
\begin{center}
\includegraphics[width=1.45\textwidth,height=0.5\textheight]{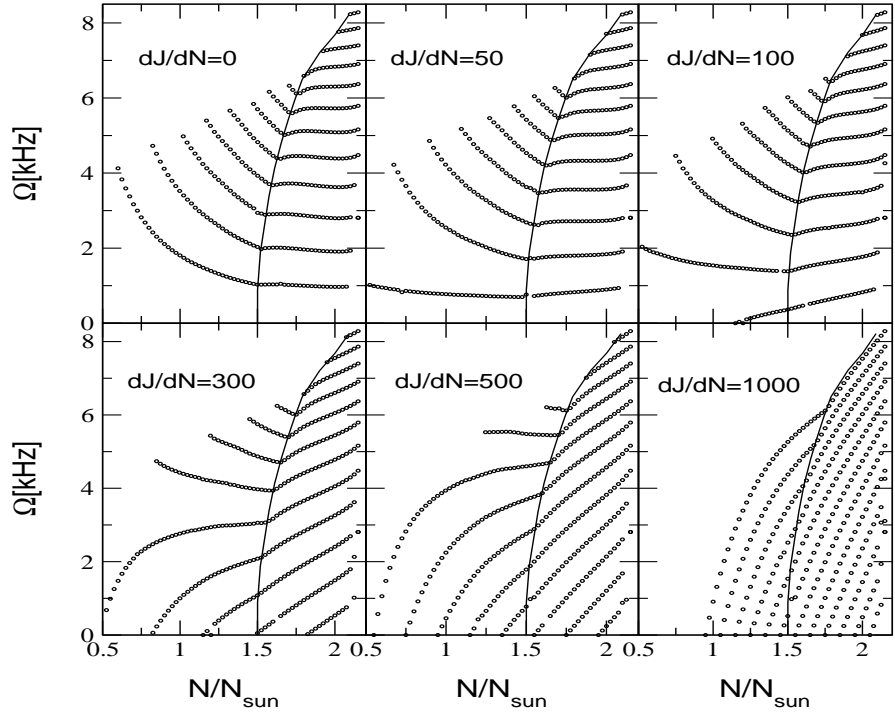}
\end{center}
\caption[]{Phase diagram for compact stars in the angular velocity - baryon 
number plane with a dividing line for quark core configurations.
Trajectories of spin evolution are given for different 
parameter values of the accretion torque $dJ/dN$ in units of 
[M$_\odot$ km$^2$ kHz/N$_\odot$] 
and different initial values $J_0$ of the angular momentum. 
The curves in the upper left panel correspond to 
$J_0$[M$_\odot$ km$^2$ kHz] $=300, 400, \dots, 1400 $
from bottom to top.}
\label{spinup}
\end{figure}

In the following we would like to explore which influence the magnitude of the 
external torque $K_{\rm ext}$ has on the pronouncedness of
the quark matter signal.  
In Fig. \ref{spinup} we show evolutionary tracks (dotted) of configurations in the 
phase diagram of Fig. \ref{phases} for different 
parameter values of the accretion torque 
and different initial values $J_0$ of the angular momentum.
\index{Accretion!pathway from canonical to ms pulsar}  

As we have discussed above, the narrow dip for $\dot \Omega$ as a quark core 
signal occurs when configurations cross the critical line during a spin-down
phase. We can quantify this criterion by introducing a minimal frequency 
$\Omega_{\rm min}$ above which spin-down occurs. 
It can be found as a solution of the equation
for the torque balance at the phase border
\begin{equation}
{d J}/{d N}= 
K_{\rm int}(N_{\rm crit}(\Omega_{\rm min}), \Omega_{\rm min}) / \dot N~.
\end{equation}

\begin{figure}[htb]
\begin{center}
\includegraphics[width=1.3\textwidth,height=0.45\textheight]{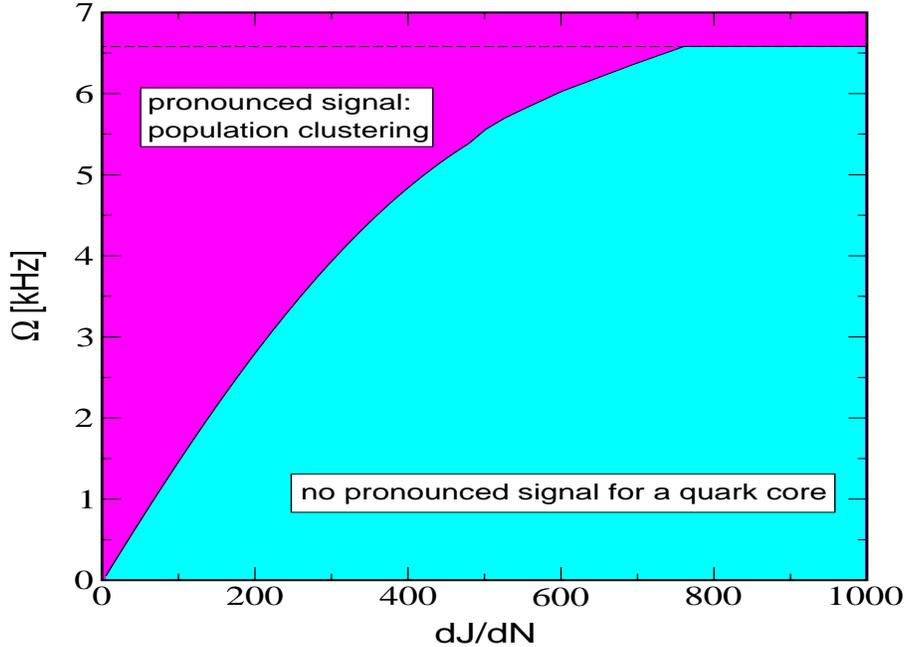}
\end{center}
\caption[]{Dividing line $\Omega_{\rm min}$ for the deconfinement signal in 
binary systems with spin-up. 
At a given rate of change of the total angular momentum 
$dJ/dN$ in units of [M$_\odot$ km$^2$ kHz/N$_\odot$] all configurations with 
spin frequency $\Omega_{\rm min}<\Omega<\Omega_{\rm max}$ have 
quark matter cores and the population clustering signal is most pronounced.
The dashed line is the maximum spin frequency 
$\Omega_{\rm max}\le \Omega_{\rm K}(N)$ for which quark core configurations 
exist.}
\label{criteri}
\end{figure}

The dependence of $\Omega_{\rm min}$ on $dJ/dN$ shown in Fig. \ref{criteri} can be
used to sample accreting compact objects from the region in which the suggested
quark matter signal should be most pronounced before making a population 
statistics.

\index{X-ray!binaries (LMZBs)}\index{QPO}
The ideal candidates for such a search program are LMXBs for which the 
discovery of strong and remarkably coherent high-frequency QPOs with the 
Rossi X-ray Timing Explorer has provided new information about the
masses and rotation frequencies of the central compact object 
\cite{klis,MLP}. 
As a strategy for the quark matter search in compact stars one should perform
a population statistics among those LMXBs 
exhibiting the QPO phenomenon which have a small $dJ/dN$
and a sufficiently large angular velocity, see Fig. \ref{criteri}. 
If, e.g., the recently discussed period clustering for Atoll- and Z-sources 
\cite{fridonew,bildsten} 
will correspond to objects in a narrow region of masses, this could be 
interpreted as a signal for the deconfinement transition 
to be associated with a fragment of the critical line in the phase diagram for 
rotating compact stars \cite{accmag}.

\section{Signal for deconfinement in LMXBs}
\index{Phase transition!deconfinement!signal}
\subsection{Spin-up trajectories for accretors}

We consider the spin evolution of a compact star under mass accretion
from a low-mass companion star as a sequence of stationary states of
configurations (points) in the phase diagram spanned by $\Omega$ and
$N$.  The process is governed by the change in angular momentum of the
star

\begin{equation} 
\label{djdt} 
\frac{d}{dt} (I(N,\Omega)~ \Omega)= K_{\rm ext}~,
\end{equation}
where
\begin{equation}
K_{\rm ext}= \sqrt{G M \dot M^2 r_0}- N_{\rm out}
\label{kex} 
\end{equation} 
is the external torque due to both the specific angular
momentum transfered by the accreting plasma and the magnetic plus
viscous stress given by $N_{\rm out}=\kappa \mu^2 r_c^{-3}$, 
$\kappa=1/3$ \cite{lipunov}. For a star with radius
$R$ and magnetic field strength $B$, the magnetic moment is given by
$\mu=R^3~B$. The co-rotating radius
$r_c=\left(GM/\Omega^2\right)^{1/3}$ is very large ($r_c\gg r_0$)
for slow rotators.
The inner radius of the accretion disc is
\[ 
r_0 \approx \left\{
\begin{array}{cc} 
R~,&\mu < \mu_c \\ 
0.52~r_A~,&\mu \geq \mu_c 
\end{array} 
\right.
\] 
where $\mu_c$ is that value of the magnetic moment of the star for
which the disc would touch the star surface.  The characteristic
Alfv\'en radius for spherical accretion with the rate $\dot M=m \dot N$
is $r_A=\left(2\mu^{-4} G M \dot M^2\right)^{-1/7}$.  Since we are
interested in the case of fast rotation for which the spin-up torque
due to the accreting plasma in Eq.  (\ref{kex}) is partly compensated
by $N_{\rm out}$, eventually leading to a saturation of the spin-up, we
neglect the spin-up torque in $N_{\rm out}$ which can be important only
for slow rotators \cite{gl},

From Eqs.  (\ref{djdt}), (\ref{kex}) one can obtain the first order
differential equation for the evolution of angular velocity

\begin{equation} 
\label{odoto} 
\frac{d \Omega}{d t}= 
\frac{K_{\rm ext}(N,\Omega)- K_{\rm int}(N,\Omega)} 
{I(N,\Omega) + {\Omega}({\partial I(N,\Omega)}/{\partial \Omega})_{N}}~, 
\end{equation} 
where 
\begin{equation}
\label{kint}
K_{\rm int}(N,\Omega)=\Omega\dot N 
({\partial I(N,\Omega)}/{\partial N})_{\Omega}~. 
\end{equation}
 
\begin{figure}[bht]
\centerline{\includegraphics[width=1.05\textwidth,clip=0]{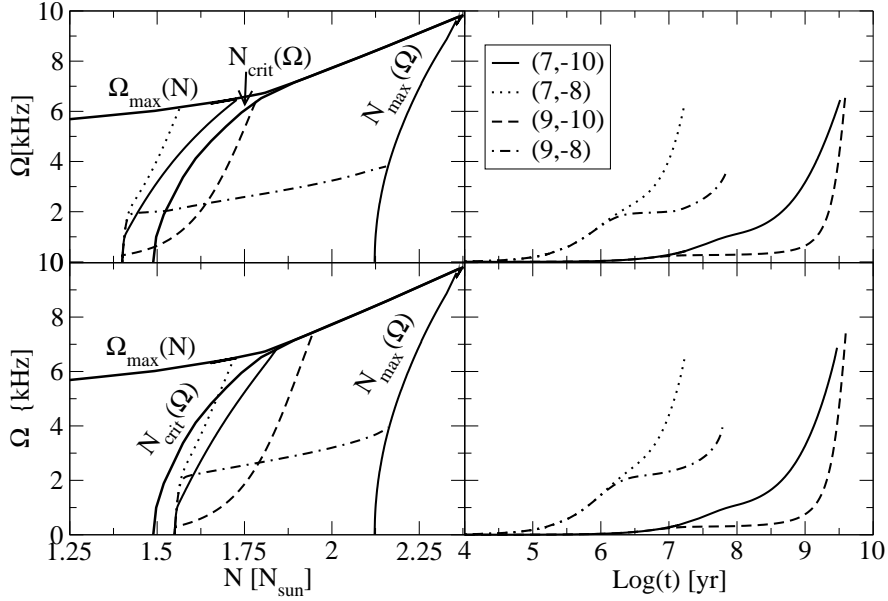}}
\caption{Spin evolution of an accreting compact star
for different decay times of the magnetic field and different accretion
rates.  Upper panels:  initial configuration with $N(0)=1.4~N_\odot$;
Lower panels:  $N(0)=1.55~N_\odot$; $\Omega(0)=1$ Hz in both cases.
The numbers in the legend box stand for ($\log (\tau_B [{\rm
yr}])$,~$\log (\dot N [N_\odot/{\rm yr}]$).  For instance (9,-8)
denotes $\tau_B=10^9$ yr and $\dot N=10^{-8}N_\odot/$ yr.
\label{fig:spinning}}
\end{figure}

Solutions of (\ref{odoto}) are trajectories in the $\Omega - N$ plane
describing the spin evolution of accreting compact stars, see Fig.
\ref{fig:spinning}.  Since $I(N,\Omega)$ exhibits characteristic
functional dependences \cite{phdiag} at the deconfinement phase
transition line $N_{\rm crit}(\Omega)$ we expect observable
consequences in the $\dot P - P$ plane when this line is crossed.

In our model calculations we assume that both the mass accretion and
the angular momentum transfer processes are slow enough to justify the
assumption of quasistationary rigid rotation without convection.  The
moment of inertia of the rotating star can be defined as $I(N,\Omega)=
J(N,\Omega)/\Omega~$, where $J(N,\Omega)$ is the angular momentum of
the star.  For a more detailed description of the method and analytic
results we refer to \cite{cgpb} and the works of
\cite{hartle,thorne}, as well as \cite{chubarian,sedrakian}.

The time dependence of the baryon number for the constant accreting
rate $\dot N$ is given by
\begin{equation} 
N(t)=N(t_0)+ (t-t_0)\dot N~.
\end{equation} 
For the magnetic field of the accretors we consider the exponential 
decay \cite{heuvel}
\begin{equation} 
B(t)=[B(0) - B_{\infty}]\exp(-t/\tau_B)+ B_{\infty}~.
\end{equation} 
We solve the equation for the spin-up evolution (\ref{odoto}) of the
accreting star for decay times $10^7\le \tau_B {\rm [yr]} \le 10^9$ and
initial magnetic fields in the range $0.2 \leq B(0){\rm [TG]}\leq 4.0
$.  The remnant magnetic field is chosen to be
$B_\infty=10^{-4}$TG\footnote[1]{1 TG= $10^{12}$ G} \cite{page}.

At high rotation frequency, both the angular momentum transfer from
accreting matter and the influence of magnetic fields can be small
enough to let the evolution of angular velocity be determined by the
dependence of the moment of inertia on the baryon number, i.e.  on the
total mass.  This case is similar to the one with negligible magnetic
field considered in \cite{cgpb,shapirov,colpi,colpiv} where $\mu \leq \mu_c$
in Eq.  (\ref{odoto}), so that only the so called internal torque term
(\ref{kint}) remains.

In Fig.  \ref{fig:spinning} we show evolutionary tracks of accretors in
phase diagrams (left panels) and show the corresponding spin evolution
$\Omega(t)$ (right panels).  In the lower panels, the paths for
possible spin-up evolution are shown for accretor models initially
having a quark matter core ($N(0)= 1.55~N_\odot, ~\Omega(0)=1$ Hz).
The upper panels show evolution of a hybrid star without a quark matter
core in the initial state ($N(0)= 1.4~N_\odot,~\Omega(0)=1$ Hz),
containing quarks only in mixed phase.  We assume a value of $\dot N$
corresponding to observations made on LMXBs, which are divided into Z
sources with $\dot N \sim 10^{-8} N_\odot/$yr and A(toll) sources
accreting at rates $\dot N \sim 10^{-10} N_\odot/$yr
\cite{klis,heuvel,gw00}.

For the case of a small magnetic field decay time $\tau_B=10^7$ yr
(solid and dotted lines in Fig.\ref{fig:spinning}) the spin-up
evolution of the star cannot be stopped by the magnetic braking term so
that the maximal frequency consistent $\Omega_{\rm max}(N)$ with stationary
rotation can be reached regardless whether the star did initially have
a pure quark matter core or not.

For long lived magnetic fields ($\tau_B=10^9$ yr, dashed and dot-dashed
lines in Fig.~\ref{fig:spinning}) the spin-up evolution deviates from
the monotonous behaviour of the previous case and shows a tendency to
saturate.  At a high accretion rate \index{Accretion!rate}
(dot-dashed lines) the mass load
onto the star can be sufficient to transform it to a black hole before
the maximum frequency could be reached whereas at low accretion (dashed
lines) the star spins up to the Kepler frequency limit.

\subsection{Waiting time and population clustering}
\index{Neutron star!rotating!population clustering}
\index{Neutron star!rotating!waiting time}
The question arises whether there is any characteristic feature in the
spin evolution which distinguishes trajectories that traverse the
critical phase transition line from those remaining within the initial
phase.

\begin{figure}[bht]
\centerline{\includegraphics[width=1.05\textwidth]{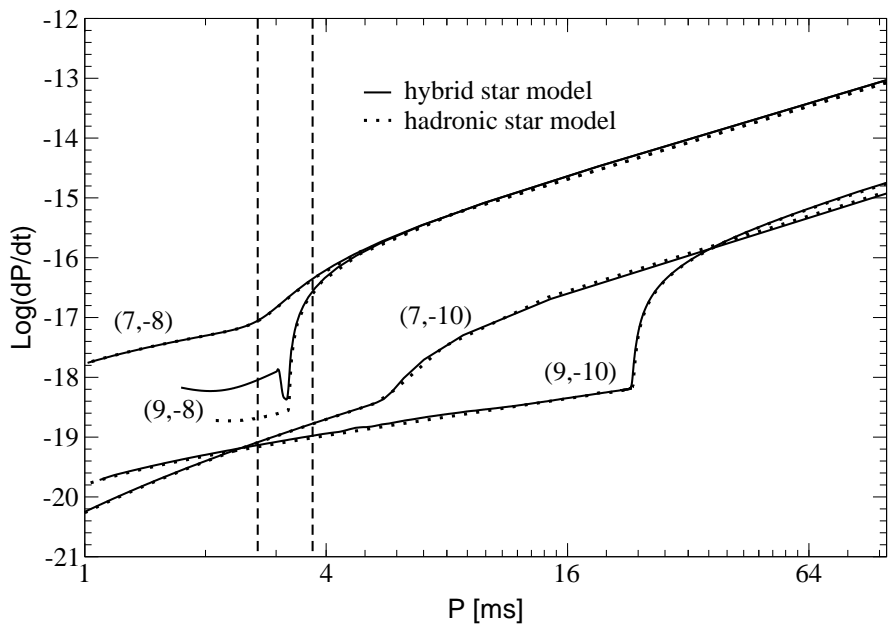}}
\caption{Spin-up evolution of accreting compact stars
in the $\dot P-P$ diagram.  For labels and initial values see to Fig.
\protect\ref{fig:spinning}.  The region of the vertical dashed lines
corresponds to the clustering of periods observed for LMXBs with QPOs.
A dip (waiting point) occurs at the deconfinement transition for
parameters which correspond to Z sources ($\dot N=10^{-8} N_\odot$/yr)
with slow magnetic field decay ($\tau_B=10^9$yr).  \label{fig:PdotP}}
\end{figure}

For an accretion rate as high as $\dot N=10^{-8}~N_{\odot}/$ yr the
evolution of the spin frequency in Fig.  \ref{fig:spinning} shows a
plateau where the angular velocity remains within the narrow region
between $2.1 \le \Omega[{\rm kHz}]\le 2.3$ for the decay time
$\tau_B=10^9$ yr and between $0.4\le\Omega[{\rm kHz}]\le 0.5$ when
$\tau_B=10^7$ yr.  This plateau occurs for stars evolving into the QCS
region (upper panels) as well as for stars remaining within the QCS
region (lower panels).  This saturation of spin frequencies is mainly
related to the compensation of spin-up and spin-down torques at a level
determined by the strength of the magnetic field.  In order to perform
a more quantitative discussion of possible signals of the deconfinement
phase transition we present in Fig.  \ref{fig:PdotP} trajectories of
the spin-up evolution in the $\dot P - P$ plane for stars with
$N(0)=1.4~N_\odot$ and $\Omega(0)=1$ Hz in the initial state; the four
sets of accretion rates and magnetic field decay times coincide with
those in Fig.  \ref{fig:spinning}.

\begin{figure}[bht]
\centerline{\includegraphics[width=1.05\textwidth]{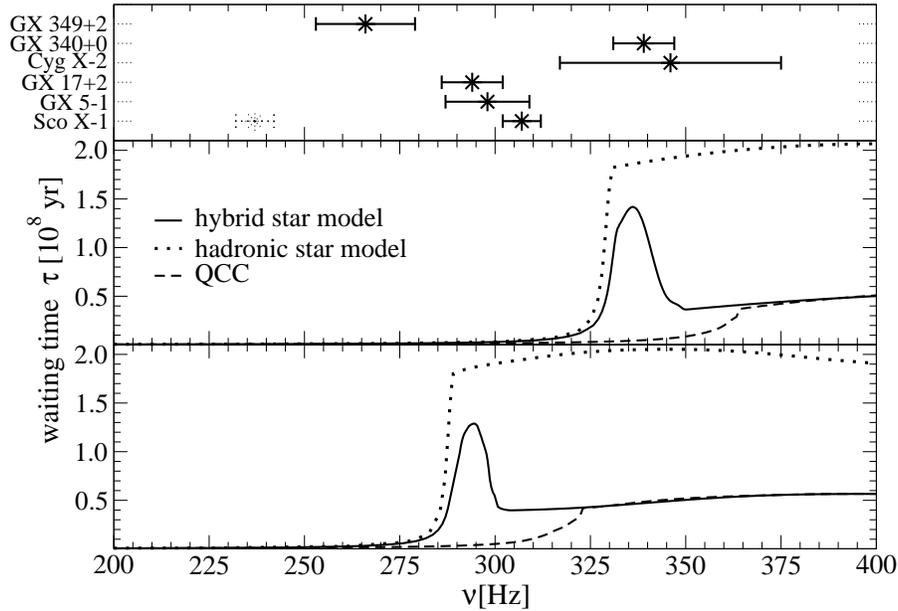}}
\caption{{\it Upper panel:}  frequency interval for
observed Z source LMXBs \protect\cite{klis}; {\it middle panel:}
waiting times $\tau=P/\dot P$ for scenario (9,-8) and initial magnetic
field $B(0)= 0.75$ TG; {\it lower panel:}  same as middle panel for
$B(0)= 0.82$ TG.  Spin evolution of a hybrid stars (solid lines) shows
a peak in the waiting time characteristic for the deconfinement
transition.  Hadronic stars (dotted lines) and QCSs (dashed lines) have
no such structure.  \label{fig:life}}
\end{figure}

When we compare the results for the above hybrid star model (solid
lines) with those of a hadronic star model (quark matter part of the
hybrid model omitted; dotted lines) we observe that only in the case of
high accretion rate ($\dot N=10^{-8}N_\odot$/yr, e.g.  for Z sources)
and long-lived magnetic field ($\tau_B=10^9$yr) there is a significant
difference in the behaviour of the period derivatives.  The evolution
of a star with deconfinement phase transition shows a dip in the period
derivative in a narrow region of spin periods.  This feature
corresponds to a plateau in the spin evolution which can be quantified
by the {\it waiting time} $\tau=\left|P/\dot P\right|=\Omega/
\dot\Omega$.  In Fig.  \ref{fig:life} (lower and middle panels) we
present this {waiting time} in dependence on the rotation frequencies
$\nu=\Omega/(2\pi)$ for the relevant case labeled (9,-8) in Figs.
\ref{fig:spinning},\ref{fig:PdotP}.  The comparison of the trajectory
for a hybrid star surviving the phase trasition during the evolution
(solid line) with those of a star evolving within the hadronic and the
QCS domains (dotted line and dashed lines, respectively), demonstrates
that an enhancement of the waiting time in a narrow region of
frequencies is a characteristic indicator for a deconfinement
transition in the accreting compact star.

The position of this peak in the waiting time depends on the initial
value of magnetic field, see Fig.  \ref{fig:life}.  In the middle and
lower panels of that Figure, we show the waiting time distribution for
$B(0)= 0.75$ TG and $B(0)= 0.82$ TG, respectively.  Maxima of the
waiting time in a certain frequency region have the consequence that
the probability to observe objects there is increased ({\it population
clustering})\index{Neutron star!rotating!pupulation clustering}.  
In the upper panel of this Figure the spin frequencies
for observed Z sources in LMXBs with QPOs \cite{klis} are shown for
comparison.  In order to interprete the clustering of objects in the
frequency interval $225 \le \nu[{\rm Hz}] \le 375$ as a phenomenon
related to the increase in the waiting time, we have to chose initial
magnetic field values in the range $1.0\ge B(0)[{\rm TG}]\ge 0.6$ for
the scenario labeled (9,-8), see also the dashed lines in 
\index{Neutron star!x-ray accretors}
\index{Neutron star!x-ray accretors!spin clustering of the population}
Fig.\ref{fig:PdotP}.

The results of the previous section show that the waiting time for
accreting stars along their evolution trajectory is larger in a
hadronic configuration than in a QCS, after a time scale when the mass
load onto the star becomes significant.  This suggests that if a
hadronic star enters the QCS region, its spin evolution gets enhanced
thus depopulating the higher frequency branch of its trajectory in the
$\Omega - N$ plane.

\begin{figure}[bht]
\centerline{\includegraphics[width=1.05\textwidth]{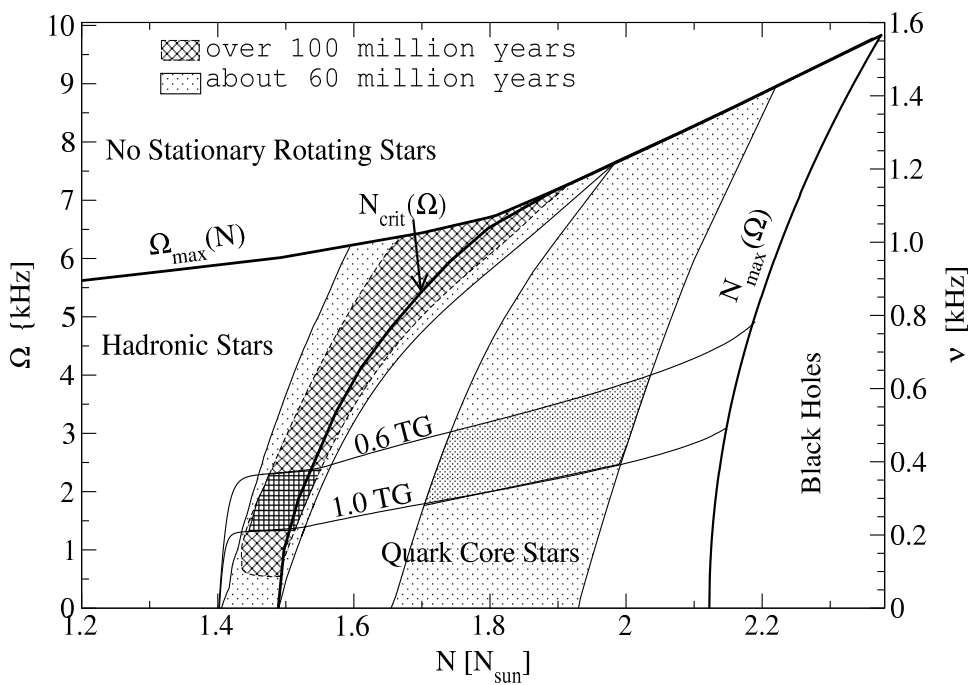}}
\caption{Regions of {waiting times} in the phase
diagram for compact hybrid stars for the (9,-8) scenario.  For an
estimate of a population statistics we show the region of evolutionary
tracks when the interval of initial magnetic field values is restricted
to $0.6\leq B(0)[{\rm TG}] \leq 1.0$.  Note that the probability of
finding a compact star in the phase diagram is enhanced in the vicinity
of the critical line for the deconfinement phase transition $N_{\rm
crit}(\Omega)$ by at least a factor of two relative to all other
regions in the phase diagram.  \label{fig:ONTcntr}}
\end{figure}

In Fig.  \ref{fig:ONTcntr} we show contours of waiting time regions in
the phase diagram.  The initial baryon number is $N(0)=1.4 N_\odot$ and
the initial magnetic field is taken from the interval $0.2\leq
B(0)[{\rm TG}] \leq 4.0$ .

The region of longest waiting times \index{Neutron star!rotating!waiting time}
 is located in a narrow branch
around the phase transition border and does not depend on the evolution
scenario after the passage of the border, when the depopulation occurs
and the probability to find an accreting compact star is reduced.
Another smaller increase of the waiting time and thus a population
clustering could occur in a region where the accretor is already a QCS.
For an estimate of a population statistics we show the region of
evolutionary tracks when the values of initial magnetic field are
within $0.6\leq B(0)[{\rm TG}] \leq 1.0$ as suggested by the
observation of frequency clustering in the narrow interval $375 \geq
\nu [{\rm Hz}] \geq 225$, see Fig.  \ref{fig:life}.

As a strategy of search for QCSs we suggest to select from the LMXBs
\index{X-ray!binaries (LMXBs)} \index{QPO}
exhibiting the QPO phenomenon those accreting close to the Eddington
limit \cite{heuvel} and to determine simultaneously the spin frequency
and the mass \cite{LM00} for sufficiently many of these objects.  The
emerging statistics of accreting compact stars should then exhibit the
population clustering shown in Fig.  \ref{fig:ONTcntr} when a
deconfinement transition is possible.  If a structureless distribution
of objects in the $\Omega - N$ plane will be observed, then no firm
conclusion about quark core formation in compact stars can be made.

For the model equation of state on which the results of our present
work are based, we expect a baryon number clustering rather than a
frequency clustering to be a signal of the deconfinement transition in
the compact stars of LMXBs.  The model independent result of our study
is that a population clustering in the phase diagram for accreting
compact stars shall measure the critical line $N_{\rm crit}(\Omega)$
which separates hadronic stars from QCSs where the shape of this curve
can discriminate between different models of the nuclear EoS at high
densities.

\section{Summary and Outlook} 
 
On the example of the deconfinement transition from hadronic to quark matter 
we have demonstrated that the rotational frequency of accreting neutron stars 
is sensitive to changes of their inner structure. 
Overclustering of the population of Z-sources of LMXBs in the frequency-mass
plane ($\Omega - N$ plane) is suggested as a direct measurement of the critical
line for the deconfinement phase transition since it is correlated to a maximum
in the moment of inertia of the compact star. 

A generalization of the phase diagram method to 
other thermodynamical observables like thermal and electromagnetic 
processes is possible and will provide a systematical tool for the study of
observable consequences of the deconfinement phase transition. 
A population statistics in the space of the appropriate thermodynamic degrees
of freedom allows a direct measurement of the shape of hypersurfaces 
representing the phase border which is characteristic for the equation of 
state of superdense stellar matter.

In the present contribution we have considered the quasistationary 
evolution of accreting objects where the timescale for accreting about one 
solar mass and thus completing the deconfinement phase transition 
processes is set by the Eddington rate to be about $10^8$ yr.
Therefore this analysis concerns rather old objects.
For the cooling process, however, a characteristic timescale is the begin of 
the photon cooling era determined mainly by the relation of neutrino and 
photon emission rates. It occurs typically after $10^2 - 10^4$ yr, i.e. for 
much younger objects. Phase transition effects, changing the emission rates,
modify cooling curves for the neutron star surface temperature typically in 
this time interval, see Voskresensky's contribution to this book 
\cite{voskresensky}.

We have focussed here on the deconfinement phase transition only. 
The phase diagram method for developing strategies to investigate changes in 
the neutron star interiors is more general and can be applied to other phase 
transitions too, as e.g. the transition to superfluid nuclear matter 
\cite{hjo}, a kaon condensate \cite{schaffner} or color superconducting quark 
matter \cite{krishnav,schaefer,craig} with typical consequences for cooling and
magnetic field evolution of the neutron star. 

In order to consider the problem of magnetic field evolution in case of color
superconductivity and diquark condensation one can use the phase
diagram method to classify the magneto-hydrodynamical effects in the 
superfluid and superconducting phases, e.g. the
creation of vortices in the quark core and their influence on the
postglitch relaxation processes can be discussed \cite{blsed}.

One can also modify the phase diagram approach to the case of self bound
rotating quark stars (strange stars) \cite{bombachi}. 
The application of our method could visualize
the evolution of quark stars from normal to strange stars.

The present approach opens new perspectives for the search for an 
understanding of the occurence and properties of superdense stellar matter in
neutron star interiors.
 
\bigskip
\medskip\noindent
{\bf Acknowledgement}
This work of H.G. and G.P. has been supported by DFG under grant Nos. 
ARM 436 17/1/00 and ARM 436 17/7/00, respectively. 
D.B. and G.P. acknowledge a visiting Fellowship granted by the ECT* Trento. 
The authors thank the participants of the workshop ``Physics of Neutron Star 
Interiors'' at the ECT* Trento for many stimulating discussions. 

\end{document}